# Novel CMOS RFIC Layout Generation with Concurrent Device Placement and Fixed-Length Microstrip Routing

Tsun-Ming Tseng†, Bing Li†, Ching-Feng Yeh‡, Hsiang-Chieh Jhan‡, Zuo-Ming Tsai‡,
Mark Po-Hung Lin‡, and Ulf Schlichtmann†
†Institute for Electronic Design Automation, Technische Universität München, Germany
‡Department of Electrical Engineering & AIM-HI, National Chung Cheng University, Taiwan

## ABSTRACT

With advancing process technologies and booming IoT markets, millimeter-wave CMOS RFICs have been widely developed in recent years. Since the performance of CMOS RFICs is very sensitive to the precision of the layout, precise placement of devices and precisely matched microstrip lengths to given values have been a labor-intensive and time-consuming task, and thus become a major bottleneck for time to market. This paper introduces a progressive integer-linear-programming-based method to generate high-quality RFIC layouts satisfying very stringent routing requirements of microstrip lines, including spacing/non-crossing rules, precise length, and bend number minimization, within a given layout area. The resulting RFIC layouts excel in both performance and area with much fewer bends compared with the simulation-tuning based manual layout, while the layout generation time is significantly reduced from weeks to half an hour.

## 1 Introduction

In wireless communication systems, RFICs are key components to receive or transmit RF signals. Millimeter-wave (mm-wave) RFICs based on CMOS process technologies have become more and more popular due to cost-effective and power-efficient system-on-chip integrations [1, 2]. Although RFICs only contain a few transistors and some passive components, such as capacitors, inductors, and transmission lines, a high-quality layout is essential as the circuit performance is very sensitive to the circuit layout, in contrast to many digital and analog designs.

In order to implement transmission lines based on CMOS technologies, thin-film microstrip lines [3], as demonstrated in Figure 1(a), are commonly adopted. A microstrip line and its ground plane are usually implemented with the top metal layer and the bottom metal layer (i.e. Metal 1), respectively. Due to the shielding of the ground plane, the lossy silicon substrate does not cause signal loss to transmission lines. As the distance, $t$, between the microstrip and its ground plane is small, which is about 5µm for 90nm CMOS technologies, the coupling effect between two microstrip lines can be neglected if the distance between them is larger than $2t$ [3, 4], or 10µm for 90nm CMOS technologies. In addition to the spacing rule, any crossing between microstrip lines is not allowed. The routing of all microstrip lines must be planar.

To achieve good RF circuit performance, layout design of RFICs, especially the routing of all microstrip lines, is extremely critical. In addition to the aforementioned spacing and non-crossing rules, any increment/decrement in length or routing bends of a microstrip line may have negative impact on circuit performance [5]. Consequently, the layout design of mm-wave CMOS RFICs have been a labor-intensive and time-consuming task. Since it is very difficult for human beings to generate an exact layout of an RFIC within a restricted layout area, designers first generate a rough initial planar layout followed by very tedious iterative simulation tuning and circuit/layout refinement. Each iteration includes i) performing full-wave electromagnetic (EM) simulation; ii) resizing devices/microstrips according to the simulation results and designers' experience; iii) adjusting the respective layout. Such manual procedure requires a large number of iterations leading to a very long layout design time. Experienced designers would spend even more than two week to finish a satisfactory layout of a 94 GHz low-noise amplifier (LNA), as shown in Figure 1(b).

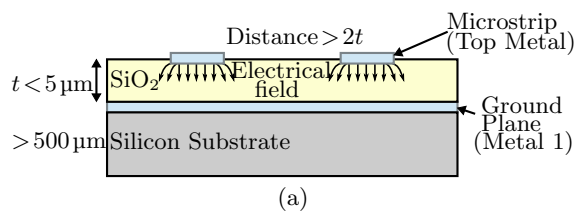
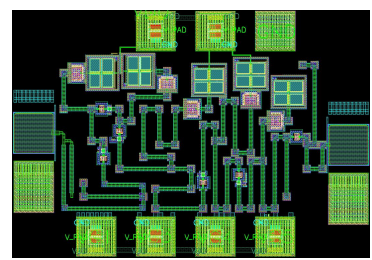

Figure 1: (a) The cross section of microstrip lines. (b) A manually designed CMOS RFIC layout (890µm×615µm) of a 94 GHz LNA with planar routing of all microstrip transimission lines.

### 1.1 Previous Work

Only few studies [6, 7] in the literature proposed automatic RFIC layout generation methods. Actuna et al. [6] focused on floorplanning, while they suggested to perform gridless maze routing afterwards. Such separation between floorplanning and routing is not suitable for CMOS RFICs with microstrip lines which need to be planar and precise. Pathak and Lim [7] presented a methodology to automatically generate RFIC layouts by iteratively performing placement and routing, and resizing circuit components to compensate performance degradation due to imprecise routing.

On length matching of wires for conventional ICs and printed circuit boards (PCBs), recent studies [8–12] tried to minimize either length difference or length-ratio difference among a set of nets during routing. Such problem formulation cannot meet the stringent routing requirements of microstrip lines in mm-wave CMOS

RFICs because the length of each microstrip line after routing must be exactly the same as the given length at the circuit design to maintain the expected RF circuit performance. Moreover, all these routing methods assume that devices are not movable, and hence fail to generate precise lengths of all microstrip lines.

## 1.2 Our Contributions

Different from the previous works [6, 7], which did not focus on length precision and bend number minimization of microstrip lines during RFIC layout generation, we propose a better layout generation methodology with an emphasis on microstrip routing optimization resulting in better performance matching before and after layout design. Our contributions are summarized below:

- We comprehensively introduce the essential routing considerations of microstrip lines, including spacing/non-crossing rules, precise length, bend smoothing, bend number minimization, and equivalent length modeling of bends;
- According to the routing considerations, we present a new problem formulation to generate the layout with precise placement and routing within a given layout area while exactly matching microstrip lengths to the given values and minimizing the number of bends;
- Based on the problem formulation, we establish a complete integer-linear-programing (ILP) model for concurrent exact placement and routing. The potential routing bends on microstrips are modeled by introducing chain points;
- In order to simplify the sophisticated ILP model, we further propose a progressive ILP-based (P-ILP) RFIC layout generation method, which consists of three different phases, considering different placement and routing abstractions.
- Compared with manual layout design, given the same or even smaller layout area, the proposed P-ILP method can reduce RFIC layout design time from weeks to half an our, and result in even better circuit performance with much fewer bends on microstrip lines.

The structure of this paper is organized as follows. Section 2 introduces microstrip routing considerations. Section 3 presents the problem formulation. Section 4 describes a general ILP model for exact placement and routing of devices and microstrips. Section 5 proposes a novel progressive RFIC layout generation method based on the ILP model. Finally, experimental results and conclusions are given in Sections 6 and 7, respectively.

## 2 Microstrip Routing Considerations

Before introducing our problem formulation, the most important routing considerations for microstrip lines should be clarified.

### 2.1 Coupling Effect

As discussed in the previous section, the spacing between microstrips and devices must be larger than two times the distance, $t$, between the layer of microstrips and the ground plane, as seen in Figure 1, to guarantee signal quality with much less coupling. To satisfy the spacing rule between any microstrips or devices, we create a bounding box around a microstrip/device which expands their horizontal and vertical dimensions by $t$ on each side, and keep any two bounding boxes from overlapping, as illustrated in Figure 2(a). We will use the expanded bounding boxes to describe the overlap constraints in our ILP model afterwards. This expansion can be easily extended to cover the case with different spacing rules between microstrips and devices.

### 2.2 Discontinuity Effect

Bends on a transmission line may cause signal loss, which is the major source of discontinuity effects [5]. To mitigate the discontinuity effect, it is essential to minimize the number of bends on microstrip lines. As the bends are sometimes unavoidable due to limited layout area, we propose to model potential microstrip bends with *chain points*, as shown in Figure 2(b). A chain point can decompose a microstrip line into two segments. Each microstrip segment has two chain points located on both ends. With chain points, each microstrip segment becomes rectangular, like a device, with two dimensions. However, a microstrip segment is more complex than an ordinary device because one of its dimensions is flexible and will be determined during microstrip routing.

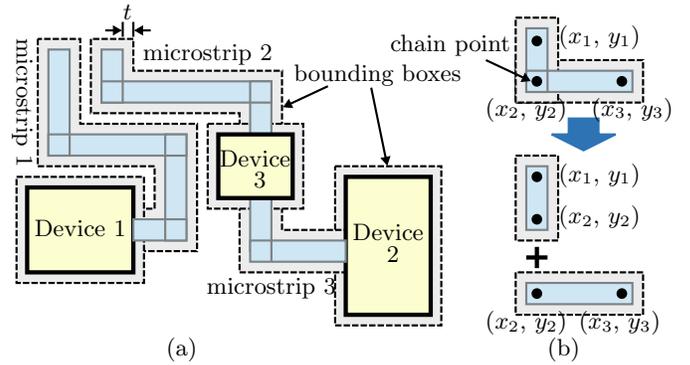

Figure 2: (a) Expanded bounding boxes of devices and microstrips for satisfying spacing rules of microstirp lines due to the coupling effect. (b) Microstrip bend modeling with chain points.

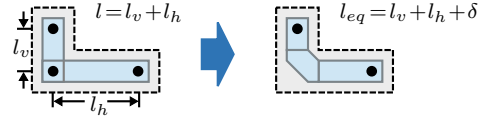

Figure 3: Bend smoothing due to discontinuity effects and equivalent length modeling.

In addition to minimizing the number bends on microstrip lines, any 90° bend must be smoothed, or replaced by a diagonal shortcut, as demonstrated in Figure 3, for reducing signal loss. Such transformation results in a different microstrip length for signal propagation, which cannot be directly represented by its geometrical length. Instead, an equivalent length change, $\delta$, must be calculated by RF simulation of the diagonal bend and comparing the signal propagation with the case through a straight microstrip. In other words, each time when a signal goes through this diagonal bend, the propagation characteristics are equivalent to the case that it goes through a straight microstrip with the equivalent length, $l_{eq} = l_v + l_h + \delta$.

Such equivalent length modeling makes our task of precise microstrip routing easier, since we only need to consider the sum of segment lengths of a microstrip line before bend smoothing, and then count the number of bends for length compensation. If there are $n$ bends, we simply add $n\delta$ to the sum of horizontal and vertical lengths to calculate the equivalent length of the microstrip line. Note that other patterns similar to the diagonal bend in Figure 3 can also be used for transmission line smoothing. The method discussed in the following sections can be adapted easily to incorporate these patterns into the proposed method.

## 3 Problem Formulation

To generate a layout for an RFIC with precise placement and routing, the input, constraints and output are detailed below.

*Input*: i) The netlist of the circuit; ii) The dimensions of the layout area; iii) The dimensions of devices; iv) The width of microstrips; v) The required distance between microstrip segments and/or devices; vi) The equivalent length compensation, $\delta$, for a smoothed bend; vii) The exact lengths of all microstrip lines.

*Constraints*: i) The equivalent lengths of microstrips should be equal to the given values; ii) No overlap exists between microstrip segments and/or devices due to the planar routing requirement; iii) Pads should be placed at the boundary of the layout area.

*Output*: A layout with a minimized number of microstrip bends.

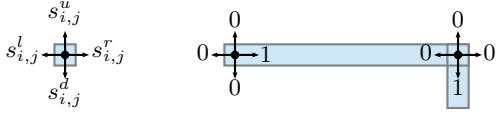

Figure 4: 0-1 variables representing segment directions at chain points.

## 4 ILP Model for Exact Placement and Routing

We first describe a general ILP model for concurrent exact placement and routing, which covers the following constraints and objectives: i) The equivalent length of a microstrip must be *equal to* the given value; ii) The bends on microstrips should be as few as possible. iii) Pads must be located at the boundary of the layout area; iv) The distance between the bounding boxes of two microstrip segments or devices must not be smaller than zero.

### 4.1 Modeling Microstrip Length

As shown in Figure 2(b), a microstrip line is decomposed into several horizontal and vertical segments. Although the 90° bends will be finally smoothed, as seen in Figure 3, the routing of a microstrip line can still be described by the corner coordinates at the 90° bends, while the length compensation value, $\delta$, will be additionally added for each bend during length calculation. Based on the chain point model, if the second segment simply follows the direction of the first segment, no real bend will be created, and the length will not be compensated by $\delta$.

Assume that there are totally $m$ microstrips in the circuit and the $i^{th}$ microstrip has $n_i$ chain points, meaning that this microstrip is formed by $n_i-1$ consecutive segments. The coordinates of these chain points are written as $(x_{i,j}, y_{i,j})$, where $j=1,...n_i$. With the coordinates of chain points, we can calculate the length $l_{i,j}$ of the $j^{th}$ microstrip segment that starts from $(x_{i,j}, y_{i,j})$ and ends at $(x_{i,j+1}, y_{i,j+1})$ as $l_{i,j}=|x_{i,j+1}-x_{i,j}|+|y_{i,j+1}-y_{i,j}|$. In this expression, there is always one term equal to 0 because the segment either spans horizontally or vertically. To linearize this representation, we need to know the relative locations of these two chain points, so that the operator of absolute value can be removed. For example, if the segment spans from left to right, we know that $x_{i,j+1} > x_{i,j}$ and $y_{i,j+1} = y_{i,j}$, so that the length of the segment can be simplified as $l_{i,j} = x_{i,j+1} - x_{i,j}$.

We represent the possible directions of the segment starting from the chain point at $(x_{i,j}, y_{i,j})$ by four directional 0-1 variables $s^u_{i,j}$, $s^d_{i,j}$, $s^l_{i,j}$ and $s^r_{i,j}$, corresponding to the up, down, left and right directions in a two-dimensional space, as illustrated in Figure 4. In this example, the directional variables for the segment from left to right are set as $s^u_{i,j} = s^d_{i,j} = s^l_{i,j} = 0$ and $s^r_{i,j} = 1$. Because a microstrip segment can take only one direction, the four 0-1 directional variables satisfy the following constraint.

$$s^u_{i,j} + s^d_{i,j} + s^l_{i,j} + s^r_{i,j} = 1, \quad i=1,...m, \quad j=1,...n_i, \quad (1)$$

where $n_i$ is the number of the chain points on the $i^{th}$ microstrip and $m$ is the number of microstrips in the circuit. Furthermore, the $(j+1)^{th}$ segment should not go back to the $j^{th}$ chain point so that the variables representing two reversed directions at two consecutive chain points should not be 1 at the same time, hence

$$s^u_{i,j} + s^d_{i,j+1} \leq 1, \quad (2)$$
$$s^d_{i,j} + s^u_{i,j+1} \leq 1, \quad (3)$$
$$s^l_{i,j} + s^r_{i,j+1} \leq 1, \quad (4)$$
$$s^r_{i,j} + s^l_{i,j+1} \leq 1. \quad (5)$$

With the four directional variables above, we can calculate the length of the $j^{th}$ segment on the $i^{th}$ wire as follows.

$$l_{i,j} = s^u_{i,j}(y_{i,j+1} - y_{i,j}) + s^d_{i,j}(y_{i,j} - y_{i,j+1}) \\ + s^l_{i,j}(x_{i,j} - x_{i,j+1}) + s^r_{i,j}(x_{i,j+1} - x_{i,j}). \quad (6)$$

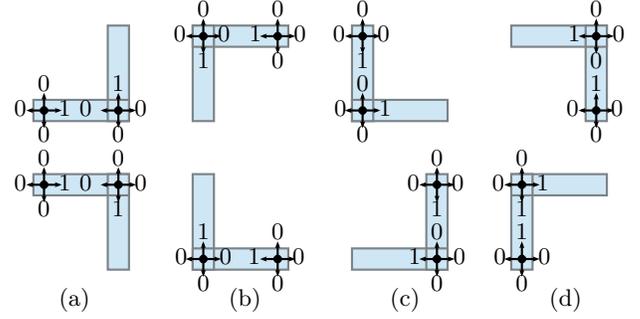

Figure 5: All kinds of bends described using the directional variables at chain points.

Each term in (6) is in the form of a multiplication of a 0-1 variable and the difference of two continuous variables, so (6) can be transformed to linear constraints according to [13]. The geometrical length of a microstrip line can thus be calculated by summing up the lengths of all segments,

$$l_{g,i} = \sum_{j=1,...n_i-1} l_{i,j}, \quad i=1,...m. \quad (7)$$

### 4.2 Modeling Bends with Chain Points

As discussed in Section 2.2, the discontinuity effect exists at each bend on the microstrip. A bend happens when two consecutive segments take different directions to span, as seen in Figure 4. To describe the condition whether a bend is really formed at a chain point in the model, we assign a new 0-1 variable $t_{i,j}$ for the chain point at $(x_{i,j}, y_{i,j})$. This variable is set to 1 if the segment starting from $(x_{i,j}, y_{i,j})$ takes a direction different from the previous segment which starts from $(x_{i,j-1}, y_{i,j-1})$.

The situations under which a bend is created are summarized in Figure 5. Only one of these situations may happen at a chain point if there is a bend, so the condition for the presence of a bend can be described as

$$s^r_{i,j-1} + s^l_{i,j-1} + s^u_{i,j} + s^d_{i,j} = 2t_{i,j,hv} + u_{i,j,hv}, \quad (8)$$
$$s^u_{i,j-1} + s^d_{i,j-1} + s^r_{i,j} + s^l_{i,j} = 2t_{i,j,vh} + u_{i,j,vh}, \quad (9)$$
$$t_{i,j} = t_{i,j,hv} + t_{i,j,vh} \leq 1, \quad (10)$$

where $t_{i,j,hv}$, $u_{i,j,hv}$, $t_{i,j,vh}$ and $u_{i,j,vh}$ are auxiliary 0-1 variables. (8) is the constraint for the cases in Figures 5(a) and (b), where $s^r_{i,j-1}$ and $s^l_{i,j-1}$ cannot be 1 at the same time according to (1), and neither can $s^u_{i,j}$ and $s^d_{i,j}$. If any of the four situations in Figures 5(a) and (b) happens, the sum on the left side of (8) is equal to 2 so that $t_{i,j,hv}$ must be set to 1. Otherwise, $t_{i,j,hv}$ must be set to 0. Similar to (8), (9) is the constraint for the cases in Figures 5(c) and (d). Combining these situations together, a bend is created when either $t_{i,j,hv} = 1$ or $t_{i,j,vh} = 1$. Therefore, the variable $t_{i,j}$ representing the presence of a bend can be constrained by (10), where $t_{i,j,hv}$ and $t_{i,j,vh}$ cannot be 1 at the same time.

The total number of real bends formed on the $i^{th}$ microstrip can thus be described as

$$n_{b,i} = \sum_{j=2,...n_i-1} t_{i,j}, \quad i=1,...m. \quad (11)$$

According to Section 2.2, a 90° bend will be replaced by a diagonal pattern in the final routing, as illustrated in Figure 3. For each bend, we need to compensate the length of the microstrip by $\delta$. Combining with the 0-1 variable $t_{i,j}$ representing the presence of a bend with the geometrical length of a microstrip defined in (7), we can write the equivalent length of the $i^{th}$ microstrip as

$$l_{eq,i} = l_{g,i} + \sum_{j=2,...n_i-1} t_{i,j}\delta. \quad (12)$$

This equivalent length must be equal to the exact length, $L_i$, of the $i^{th}$ microstrip according to the specification. Consequently,

$$l_{eq,i} = L_i \quad (13)$$

### 4.3 Modeling Connections to Devices and Pads

The two ends of a microstrip should be connected to either devices or pads. Assume that the connection point on the microstrip has the coordinate $(x_{i,j}, y_{i,j})$, which is equal to $(x_{i,1}, y_{i,1})$ when the starting chain point is connected, or $(x_{i,n_i}, y_{i,n_i})$ when the ending chain point is connected. Assume that the center of the $k^{th}$ device/pad is $(x_k, y_k)$ and a pin on it has the offset $(x_t, y_t)$ from the center. If this pin is connected to the chain point $(x_{i,j}, y_{i,j})$, the two coordinates must be the same, which can be described as

$$x_{i,j} = x_k + x_t \quad \text{and} \quad y_{i,j} = y_k + y_t. \quad (14)$$

Note that on some devices, some pins might be equivalent, so the locations of these pins can be switched in the model.

Pads are special devices that should be placed along the boundary of the layout area. Assume that we align the center $(x_k, y_k)$ of the $k^{th}$ device to the boundary and the dimensions of the layout area are $L_h$ and $L_v$. We introduce two discrete auxiliary variables $x_{k,d} \in \{0, L_h\}$ and $y_{k,d} \in \{0, L_v\}$, and two continuous auxiliary variables $0 \le x_{k,c} \le L_h$ and $0 \le y_{k,c} \le L_v$. A 0-1 variable $c_k$ is used to determine whether the pad is placed at the vertical or horizontal boundary. The constraint for pad placement can be written as

$$x_k = c_k x_{k,d} + (1-c_k) x_{k,c} \quad \text{and} \quad y_k = (1-c_k) y_{k,d} + c_k y_{k,c}. \quad (15)$$

Similar to (6), this constraint can be linearized according to [13].

### 4.4 Modeling Non-overlapping Conditions

The distance between any pair of microstrip segments or devices must satisfy the spacing rule. As shown in Figure 2(a), we model the spacing rules as non-overlapping constraints of the bounding boxes of devices/segments. If the distance between any two bounding boxes is not smaller than 0, the distance between any pair of microstrip segments or devices satisfies its spacing rule.

According to Figure 2, the coordinate of a corner of the bounding box of a segment can be calculated from the coordinate of the chain point and the width of the segment. Similarly, the corner coordinates of a bounding box of a device can be calculated from its center and its dimensions. For the convenience of expression, we explain the non-overlapping constraints with these corner coordinates directly. Assume that the upper-left corner and the lower-right corner of the $i^{th}$ block have coordinates $(x_i^l, y_i^u)$ and $(x_i^r, y_i^d)$, respectively, and the corresponding coordinates of the $j^{th}$ blocks are $(x_j^l, y_j^u)$ and $(x_j^r, y_j^d)$, respectively. If these two blocks are not overlapped, their relative locations must be one of the situations in Figures 6(b)–(e), so that the corner coordinates of the bounding boxes should meet

$$x_i^r \le x_j^l + \mathcal{M} u_{i,j,1}, \quad (16)$$
$$y_j^u \le y_i^d + \mathcal{M} u_{i,j,2}, \quad (17)$$
$$x_j^r \le x_i^l + \mathcal{M} u_{i,j,3}, \quad (18)$$
$$y_i^u \le y_j^d + \mathcal{M} u_{i,j,4}, \quad (19)$$
$$u_{i,j,1} + u_{i,j,2} + u_{i,j,3} + u_{i,j,4} \le 3, \quad (20)$$

where $u_{i,j,1}$–$u_{i,j,4}$ are auxiliary 0-1 variables, and $\mathcal{M}$ is a large constant. The constraint (20) requests that at least one variable from $u_{i,j,1}$–$u_{i,j,4}$ to be set to 0, and hence at least one of the four situations in Figure 6 is guaranteed.

### 4.5 Optimization Formulation

The objective of the optimization problem described in Section 3 is to minimize total number of bends on all microstrips. According to the discussion in previous sections, the overall ILP model for

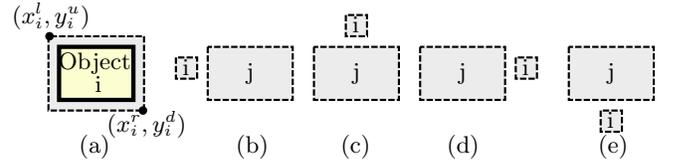

Figure 6: (a) Corner coordinates of a bounding box. (b)–(e) Non-overlapping situations.

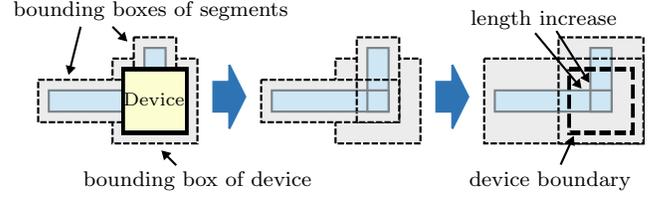

Figure 8: Space reservation for blurred devices.

exact placement and routing of an RFIC can be defined as

$$\text{Minimize:} \quad \alpha n_{b,max} + \beta \sum_{i=1,\ldots m} n_{b,i}, \quad (21)$$

$$\text{Subject to:} \quad (1)-(20). \quad (22)$$

In the above optimization problem, the ILP solver needs to search the complete layout area to determine where the devices and chain points on microstrips should be located. The large search space and variable numbers make the problem very difficult to solve, especially to meet all the exact length specifications.

## 5 Progressive ILP-based RFIC Layout Generation

Although the accurate ILP model presented in the previous section can generate an exact layout, including placement and routing, the runtime is not acceptable. To efficiently solve the problem, we apply simplified versions of this model in three different phases based on different levels of layout abstraction, and limit the solution space, respectively. Consequently, an optimized RFIC layout can be automatically generated with reasonable runtime.

The proposed three phases include i) planar microstrip routing with blurred devices, ii) device visualization and overlap fixing, and iii) iterative layout refinement with device rotation and deletion/insertion of chain points. Figure 7 demonstrates the overall progressive ILP-based RFIC layout generation flow with the snapshot resulting from each phase.

### 5.1 Planar Microstrip Routing with Blurred Devices

We first generate the planar routing of all microstrip lines with blurred devices by excluding detailed device geometries in the ILP model described in Section 4. Although devices are not directly considered in this phase, their dimensions are integrated into the model by enlarging the space between microstrips and/or devices. The concept and idea of this space reservation is illustrated in Figure 8. With these aggressively expanded bounding boxes, sufficient space can be reserved for blurred devices during planar microstrip routing.

After blurring devices, different microstrip lines are directly connected, so that their lengths are also increased according to the dimensions of the devices, as shown in Figure 8. These increased lengths should be added to the specified microstrip length $L_i$ in (13). Assume that the increased lengths at the starting device and the ending device of the $i^{th}$ microstrip are $L_{s,i}$ and $L_{e,i}$, respectively. The length requirement of this microstrip becomes

$$L_{gr,i} = L_i + L_{s,i} + L_{e,i} \quad (23)$$

However, during planar microstrip routing, we might expand the bounding boxes of microstrip segments too aggressively, and hence the lengths of microstrips may not be completely satisfied. We alleviate such problem by minimizing the maximum unmatched length on microstrips and the total unmatched length of all of them. For this purpose, we introduce the variable $l_{u,i}$ to represent the upper bound of the under-compensated or over-compensated length of the $i^{th}$ microstrip, so that $l_{u,i} \geq |L_{gr,i} - l_{eq,i}|$, where $L_{gr,i}$ is the specified length for the $i^{th}$ microstrip in global routing defined in (23), and $l_{eq,i}$ is the equivalent length of the microstrip defined in (12). This constraint can also be expressed as

$$l_{u,i} \geq L_{gr,i} - l_{eq,i} \quad \text{and} \quad l_{u,i} \geq l_{eq,i} - L_{gr,i}, \quad i=1,\ldots m. \quad (24)$$

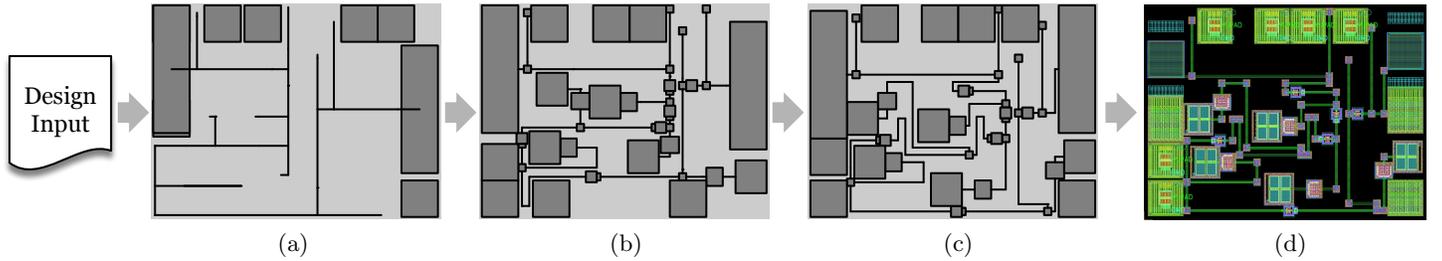

Figure 7: The snapshot resulting from each phase of the proposed progressive RFIC layout generation flow based on the same 94 GHz LNA, as seen in Figure 1(b). (a) Planar microstrip routing with blurred devices. (b) Device visualization and overlap fixing. (c) Iterative layout refinement with device rotation and deletion/insertion of chain points. (d) The resulting layout (800µm×600µm).

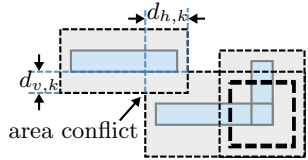

Figure 9: Overlap between bounding boxes.

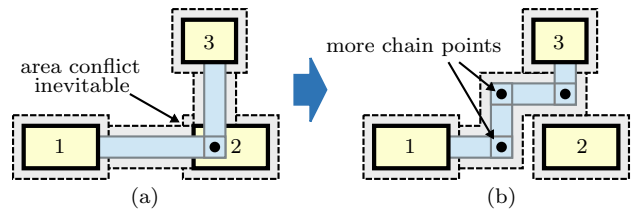

Figure 10: Chain point insertion during layout refinement. (a) No feasible solution with insufficient chain points. (b) A feasible solution with sufficient chain points.

The maximum of all these bounds is modeled with the variable $l_{u,max}$ and constrained as

$$l_{u,max} \geq l_{u,i}, \quad i=1,\ldots m. \quad (25)$$

In addition, the bounding box expansion in Figure 8 also reserves space aggressively and therefore area overlap cannot be completely avoided. In this phase, we allow some overlap between the expanded bounding boxes. These overlap areas are penalized in the optimization problem so that the solver still tries to reduce them. Consider the example in Figure 9. We represent the dimensions of the $k^{th}$ overlap area with variables $d_{h,k}$ and $d_{v,k}$, respectively. Then the sum of all these overlap dimensions is minimized together with unmatched microstrip lengths.

Similar to the optimization problem in (21)–(22), the ILP problem for planar microstrip routing with blurred devices is defined as follows.

$$\text{Minimize:} \quad \alpha n_{b,max} + \beta \sum_{i=1,\ldots m} n_{b,i} + \gamma l_{u,max}$$
$$+ \zeta \sum_{i=1,\ldots m} l_{u,i} + \eta \sum_{k \in \mathcal{K}} (d_{h,k} + d_{v,k}), \quad (26)$$

$$\text{Subject to:} \quad (1)-(20) \text{ except } (14), \quad (27)$$
$$(23)-(25), \quad (28)$$

where $\mathcal{K}$ is the set of indexes of all possible overlapping areas created by comparing all the bounding boxes in the current phase. In this phase, we set the number of chain points on a microstrip $n_i$ to a given number to reduce model complexity. If this setting is too restrictive, more chain points are inserted in the later refinement iterations.

## 5.2 Device Visualization and Overlap Fixing

After planar microstrip routing, the blurred devices are visualized again for overlap fixing and routing refinement. The locations of all blurred devices can be obtained according to the starting and ending points of their connected microstrips. Once the blurred devices are visualized at the corresponding locations, we want to fix device overlap and release the unused space which was previously reserved along microstrip segments. We solve the optimization problem (26)–(28) again with the new area constraints. The only difference is that we include the constraint (14) into the ILP model for considering device geometries and pin locations in this phase.

In order to reduce the solution space, the locations of chain points on microstrips are not allowed to freely move across the whole layout area since the planar microstrip routing in the previous phase has determined the topology of all microstrip segments. Therefore, chain points and device locations are confined to areas with size equal to $\tau_d$ centering at their current coordinates.

## 5.3 Iterative Layout Refinement with Chain Point Deletion/Insertion and Device Rotation

In this phase, the optimization problem (26)–(28) is further solved several times, taking the result of the second phase as the initial solution. To achieve better solutions in each iteration, we refine the ILP model of the optimization problem by i) deleting chain points on microstrips, ii) inserting chain points on microstrips, and iii) rotating devices. Similar to the second phase, chain points and devices are also confined in their local areas in this phase.

As chain points are used to represent where microstrips can change direction, they are virtual and used only for modeling. Each time after solving the ILP problem, there might be no bends formed at some chain points. In other words, the two microstrip segments chained by such a chain point have the same direction. These chain points without bends can be removed to reduce solution space during iterative layout refinement.

Sometimes the reduced number of chain points may affect the routing of microstrips. For example, in Figure 10(a), with only one chain point it is impossible to create a microstrip connecting device 1 and device 3 without overlapping device 2. To solve this problem, we can either change the orientations of some devices or insert some chain points on a microstrip. The insertion of chain points and rotation of devices enables the solver to generate a valid routing to circumvent overlap, as shown in Figure 10(b), or to achieve even more precise microstrip routing with fewer bends.

## 6 Experimental Results

The proposed framework was implemented with the C++ programming language, and executed on a computer with a 2.67 GHz CPU and 12 GB memory. We employed Gurobi Optimizer [14] as our ILP solver. We tested the proposed P-ILP method on three RF circuits, as shown in Table 1, and compared the resulting layout quality of each circuit with the respective manually designed layouts given by RFIC designers. For each RF circuit, we applied two different area settings. In the first setting, the dimensions of the area are exactly the same as those of the manually designed layout. In the second setting, we applied a smaller area with a similar aspect ratio for stress testing of our approach when the layout density is even higher.

Table 1: Comparison of maximum bend numbers on a microstrip line, total bend numbers on all microstrip lines, and runtime for the RFIC layouts resulting from manual design ("Manual") and the proposed progressive ILP-based approach ("P-ILP").

| Circuit | # of microstrips | # of devices | Area (µm×µm) | Max. bend number Manual | Max. bend number P-ILP | Total bend number Manual | Total bend number P-ILP | Runtime Manual | Runtime P-ILP |
|---|---|---|---|---|---|---|---|---|---|
| 94 GHz LNA | 25 | 34 | 890×615 | 9 | 4 | 59 | 22 | >2 weeks | 18m05s |
| | | | 845×580 | n/a | 5 | n/a | 29 | n/a | 28m13s |
| 60 GHz Buffer | 14 | 26 | 595×850 | 4 | 3 | 27 | 7 | >1 week | 04m22s |
| | | | 505×720 | n/a | 3 | n/a | 13 | n/a | 19m20s |
| 60 GHz LNA | 19 | 28 | 600×855 | 4 | 2 | 31 | 10 | >1 week | 06m17s |
| | | | 570×810 | n/a | 5 | n/a | 18 | n/a | 07m12s |

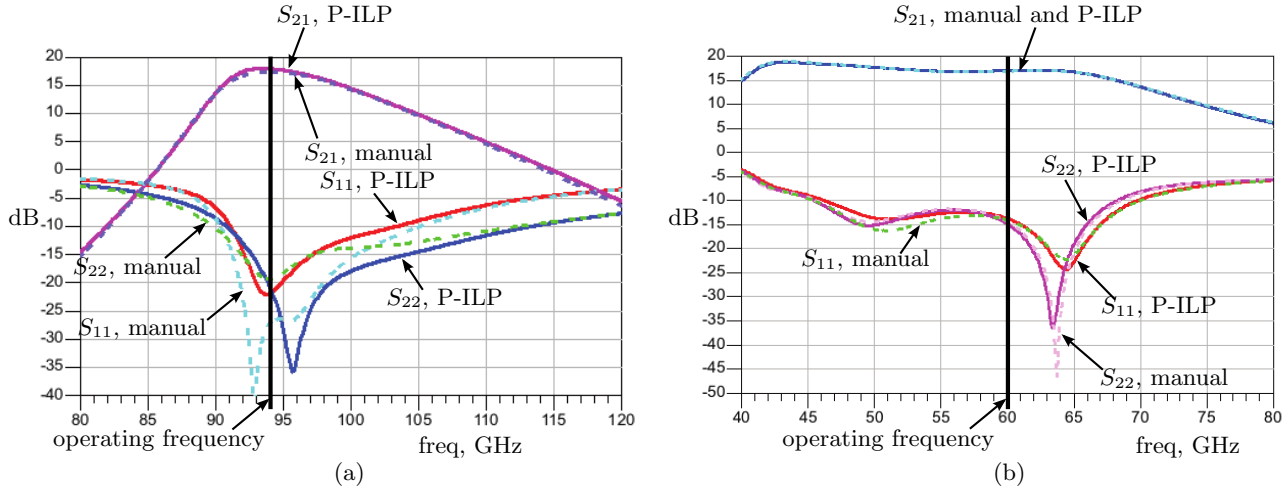

Figure 11: RF simulation results. (a) 94 GHz LNA, manual design with area 890µm×615µm and P-ILP with area 800µm×600µm. (b) 60 GHz Buffer, manual design with area 595µm×850µm and P-ILP with area 500µm×800µm.

According to the results in Table 1, when applying the same area setting, our P-ILP method results in even smaller maximum bend number on a microstrip line and much fewer total bends on all microstrip lines in a circuit, compared with the manually designed layouts. The layout design of each circuit for an experienced designer requires at least one week due to tedious polygon pushing and simulation tuning. Our P-ILP method takes at most half an hour to accomplish the layout of an RF circuit with even better layout quality. When applying the smaller area setting, our approach can still generate a feasible layout for each circuit with much fewer bends within acceptable running time. Consequently, our P-ILP method is very effective and efficient.

In addition to the number of bends on microstrip lines, to verify RF performances, we further simulated two of the circuits using Agilent Advanced Design System (ADS). The performances of the manual layout and the automatically generated layout of 94 GHz LNA and 60 GHz Buffer are shown in Figure 11. The major performance of these circuits is the gain, $S_{21}$, from port 1 to port 2 at the operating frequencies as highlighted in Figure 11. In the LNA circuit, the gain values of P-ILP and manual design are 17.912 dB and 17.196 dB, respectively. In the Buffer circuit, the gain values are 16.998 dB and 16.791 dB, respectively. Besides gain, the return loss measurements at port 1 and port 2 are denoted by $S_{11}$ and $S_{22}$ in Figure 11, respectively. For port return loss, manual design and P-ILP each have an advantage in some performances in the regions around the operating frequencies. In summary, our P-ILP method outperforms the manual design in terms of gain, while achieving comparable quality with respect to port return loss.

Therefore, we can conclude that the proposed method consistently excels in circuit performance, chip area, and execution time.

## 7 Conclusion

In this paper, we propose to generate high-quality layout for mm-wave RFICs using an efficient design automation method. In such circuits, microstrip lines must have given lengths in the routing to maintain circuit performances. In addition, bends on microstrips should be reduced as much as possible. We model this layout generation task as an ILP problem and solve it in several phases with simplified models. Experiments show that the proposed method can generate a valid layout efficiently, while circuit performance resulting from the automatically generated layout is consistently better than manual design.